%% file: mishar2.tex
\documentclass [
aps,
12pt,
final,
notitlepage,
oneside,
onecolumn,
nobibnotes,
nofootinbib,
superscriptaddress,
showpacs,
centertags,
koi8-r]
{revtex4}

\begin {document}
\selectlanguage{english}

\input abstract_eng.tex

\graphicspath{{./eps/}{./}}

\section{Introduction}

The inclusive spectra of secondaries (pions, kaons, $p$ and $\bar{p}$)
measured in lepton--nucleus ($lA$) deep-inelastic scattering (DIS)
\cite{Aira1,Aira2,Ashman} become more and more soft with the increase
of atomic weight of the target. In this paper we consider the
$A$ dependences of these spectra using the same method as in our
previous paper \cite{Berd1}.

In the case of DIS only some part of the projectile lepton energy
($\nu=E_l-E_l^{\prime}$) is transfered to the target.
Thus, one can consider the virtual photon as the projectile to draw
an analogy between the DIS and hadroproduction processes. It is
convenient to analyze DIS data in terms of energy fractions of the
virtual photon carried by the produced hadron. Therefore in this
paper we will use the variable:
 \begin{equation}
 \label{eq:z}
 z=\frac{E_h}{\nu},
 \end{equation}
where $E_h$ is the energy of the produced hadron in the lab. frame.

At high energies in the case of nucleon target the maximum value of
$z$ is close to unity ($z_{\rm max}\to1$). In the case of nuclear
target the situation is more complicated because there are many
different contributions \cite{Kop1,Accardi} from the final-state
interactions with nuclear matter which decrease the spectra at large
$z$. On the other hand, the processes which leads to the so-called
cumulative effect \cite{Bald,Bay,Fra} increase the boundaries of the
available $z$ region\footnote{The last processes have rather small cross 
section.}. The modification of parton structure function \cite{Esk} 
should be accounted for.

In all processes on nuclear targets (except of the coherent ones) some
fraction of energy is used for nuclear disintegration. The nuclear
target is destroyed and several nucleons, as well as light nuclear
fragments, (say, $\alpha$ particles) appear in the final state.
It can be considered as a phase-space limitation for the produced
secondaries. The corresponding fraction of initial energy used on this
effect numerically is not so small (it is many times larger than the
nuclear binding energy).

The energy used for nuclear disintegration is taken from the beam or
secondary particle energy, primarely via some QCD process, discussed
in \cite{Kop1,Accardi}. After several stages this energy transforms (in part,
as a minimum) into kinetic energy of the target fragments. So the portion
of the last kinetic energy allows us to estimate the primary energy losses 
all together. On the other hand, the phase-space limitation can be 
considered as an effective decrease of the incident beam energy.

In what follows we will consider the $A$ dependences of secondary
hadron leptoproduction in terms of $z_A$:

\begin{equation}
\label{eq:zA}
z_A = \frac{E_h}{\nu - E_A},
\end{equation}
and we assume that it is possible to find the shift $E_A$ for the case
of interaction with nuclear target from the condition that the ratios
of secondary multiplicities on nuclear and nucleon ($E_{A=1} = 0$) targets
in terms of $z_A$
\begin{equation}
R_{lA/lp}(z_A) = const (z_A) \simeq 1 ,
\end{equation}
whereas the same ratios in terms of $z$
\begin{equation}
R_{lA/lp}(z) = f(z).
\end{equation}

Evidently, such rescale is reasonable only for not very small
$z$ values.

We will determine shifts $E_A$ from the experimental data and we
will compare them with several independent estimations. Such approach
allows us, in particular, to investigate the energy ($\nu$) dependence
of all nuclear effects. In conclusion we will compare our results with
theoretical calculations \cite{Kop1,Accardi}.

\section{$A$ dependence of secondary leptoproduction at \newline
large $z_A$}

The experimental results for semi-inclusive deep-inelastic scattering
on nuclei are usually presented in terms of the hadron multiplicity ratio
$R_{A/D}$, which represents the ratio of the number of hadrons of type $h$
produced per deep-inelastic scattering event on a nuclear target of mass $A$
to that from a deuterium target ($D$). The multiplicity ratio is defined as:

\begin{equation}
\label{eq:Rhm}
R_{A/D} = \frac{\frac{1}{A}\left(\frac{d\sigma}{dz}\right)_{lA\rightarrow h}}
 {\frac{1}{2}\left(\frac{d\sigma}{dz}\right)_{lD\rightarrow h}},
%R_{A/D}(z,\nu) = \frac{\left. \frac{N_h(z,\nu)}{N_l(\nu)} \right|_{A}}
%{\left. \frac{N_h(z,\nu)}{N_l(\nu)} \right|_{D}},
\end{equation}
where $\left(\frac{d\sigma}{dz}\right)_{lA\rightarrow h}$ is the yield
of semi-inclusive hadrons $h$ from the nucleus $A$ in a given $z$-bin.

In this section we are going to analyze the ratio Eq.~(\ref{eq:Rhm}) in
the following way.
Suppose one has two $z$ spectra for DIS on nucleon and nuclear targets
or on two different nuclei (light and
heavy ones). Then one can shift the spectrum that corresponds to the heavy
nucleus according to Eq. (\ref{eq:zA}) by changing $E_A$ parameter.
Assuming that nuclear effects are small
for very light nuclei, it is possible then to calculate the fraction of
the virtual photon energy spent on nuclear effects. This may be done by
calculating the ratio of the shifted spectrum to the spectrum that
corresponds to the
light nucleus. When this ratio is close to unity then the corresponding
shift
will give the absolute value of energy loss caused by the mentioned
nuclear effects.

For the purpose of our analysis we used the experimental data on DIS on
nuclei
measured by the HERMES Collaboraton \cite{Aira1,Aira2} as well as EMC data
\cite{Ashman}.

\begin{figure}[htb]
\setcaptionmargin{5mm} 
\onelinecaptionsfalse
 \includegraphics[width=.9\hsize]{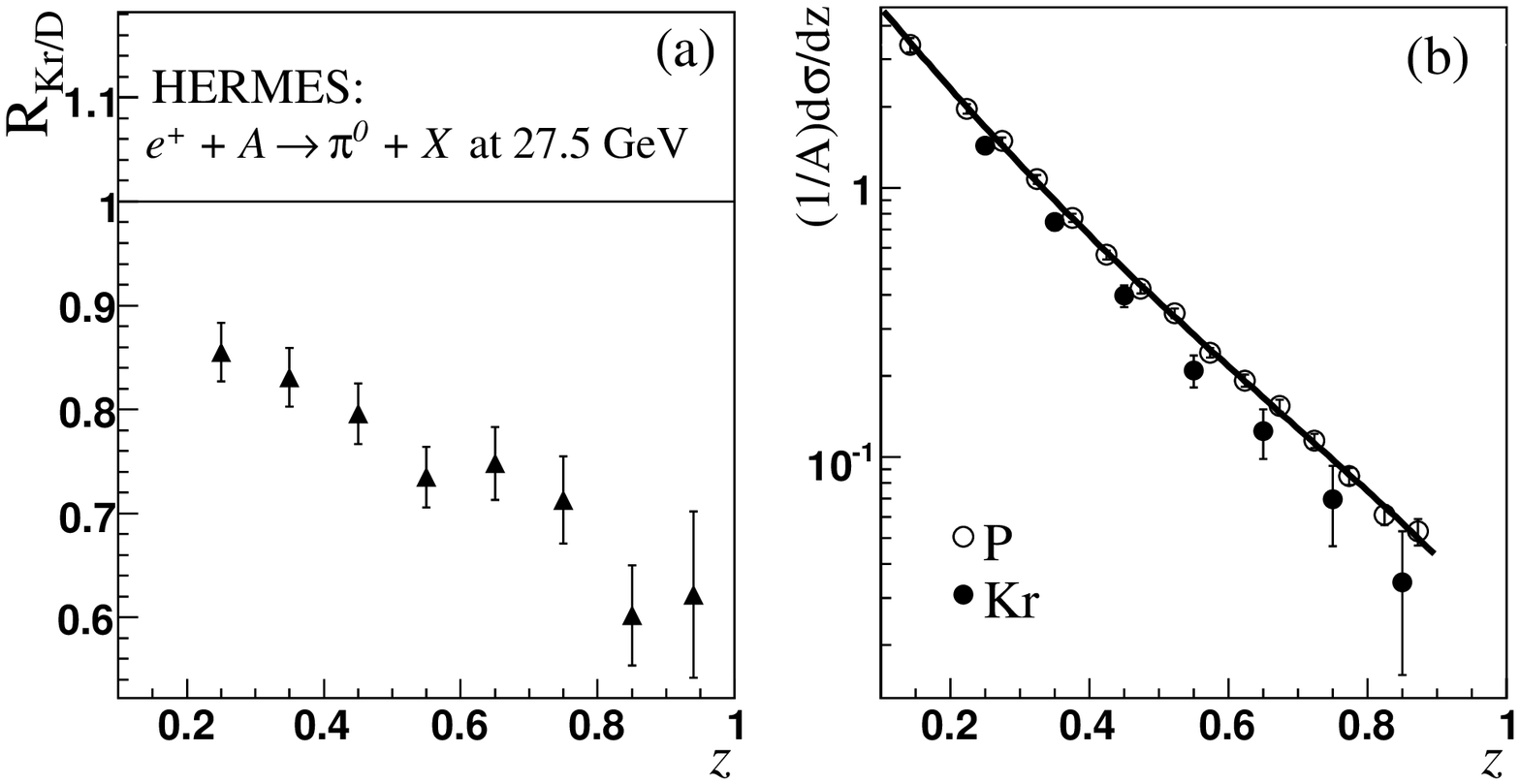}
\captionstyle{normal}
 \caption{(a) Multiplicity ratio for identified neutral pions from a Kr
target as a function of z (for $\nu>7$~GeV). (b) Neutral pion multiplicities for
hydrogen \cite{Aira2} and Kr (extracted from ratio Fig.~\ref{fig:HERMES_extract}a).}
 \label{fig:HERMES_extract}
 \end{figure}

The HERMES results for deep-inelastic $e^+$D and $e^+$Kr scattering at
27.5~GeV
are available in terms of the ratio Eq.~(\ref{eq:Rhm}) for different set
of secondaries (pions, kaons, $p$ and $\bar{p}$) \cite{Aira1}.
Neutral pion and averaged charged pion multiplicities
for DIS of positrons on hydrogen at the same energy
are published in Ref.~\cite{Aira2}. Neglecting the difference between
hydrogen and deuterium targets one can extract the multiplicity for
pions produced on a heavy target (Kr) from the multiplicity ratio and
absolute multiplicity spectrum for hydrogen.
The measured ratio for neutral pions is presented in
Fig.~\ref{fig:HERMES_extract}a.
In Fig.~\ref{fig:HERMES_extract}b one can see extracted spectrum for Kr
target (solid circles)
as well as neutral pion multiplicity for hydrogen (open circles) from
Ref.~\cite{Aira2}.
One can see evident suppression of Kr spectrum in comparison with
hydrogen one.

\begin{figure}[htb]
\setcaptionmargin{5mm} 
\onelinecaptionsfalse
\includegraphics[width=.9\hsize]{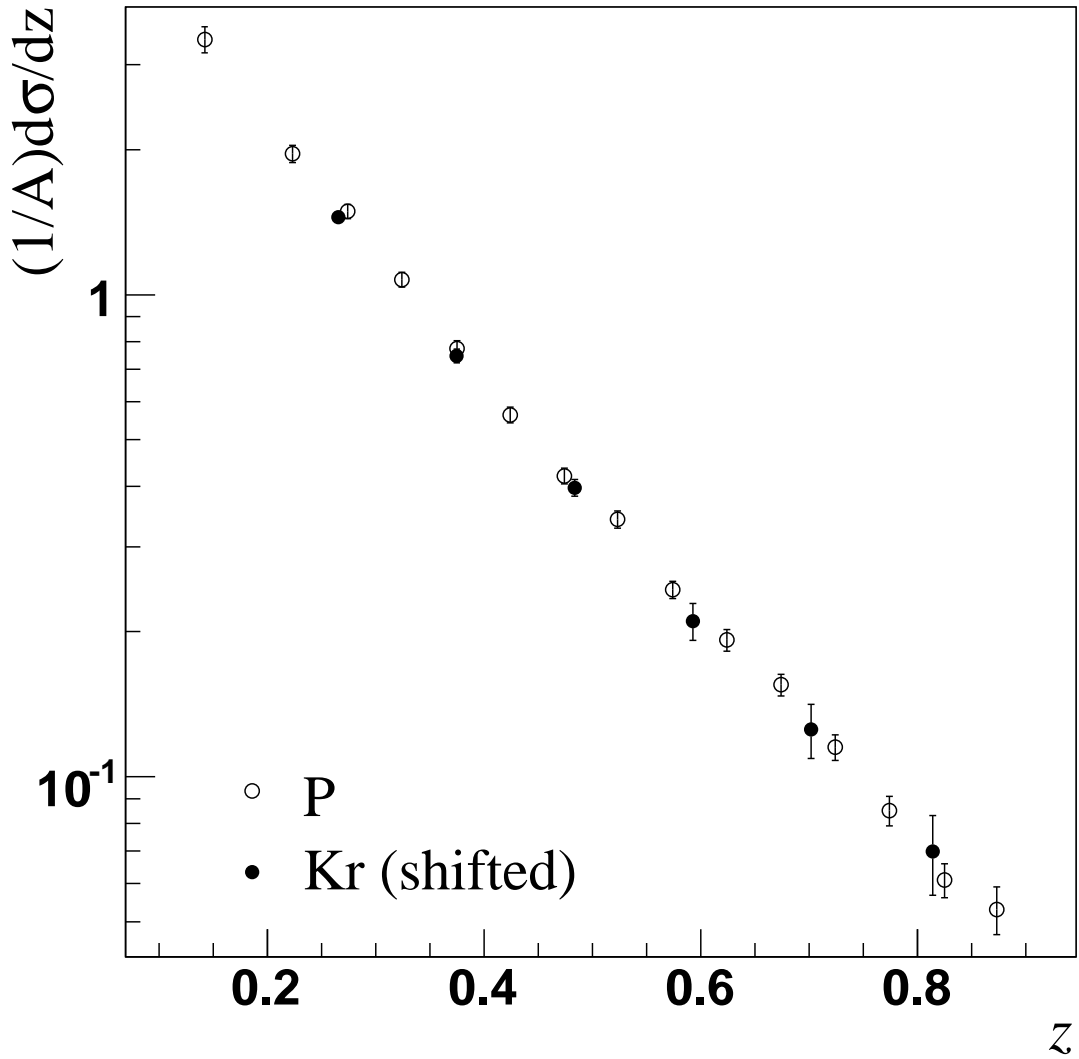}
\captionstyle{normal}
 \caption{The same as in Fig.~\ref{fig:HERMES_extract}b, but the
spectrum for
Kr target was shifted according to Eq.~(\ref{eq:xF}) with $E_A=1.0$~GeV.}
 \label{fig:HERMES_shift}
 \end{figure}

Having two multiplicity spectra for $p$ and Kr targets one can analyze them
in the way described above. Fig.~\ref{fig:HERMES_shift} represents the
shifted Kr spectrum with $E_A=1$~GeV (solid circles) as well as spectrum for hydrogen
measured by
HERMES (open circles).

\begin{figure}[htb]
\setcaptionmargin{5mm} 
\onelinecaptionsfalse
\includegraphics[width=.9\hsize]{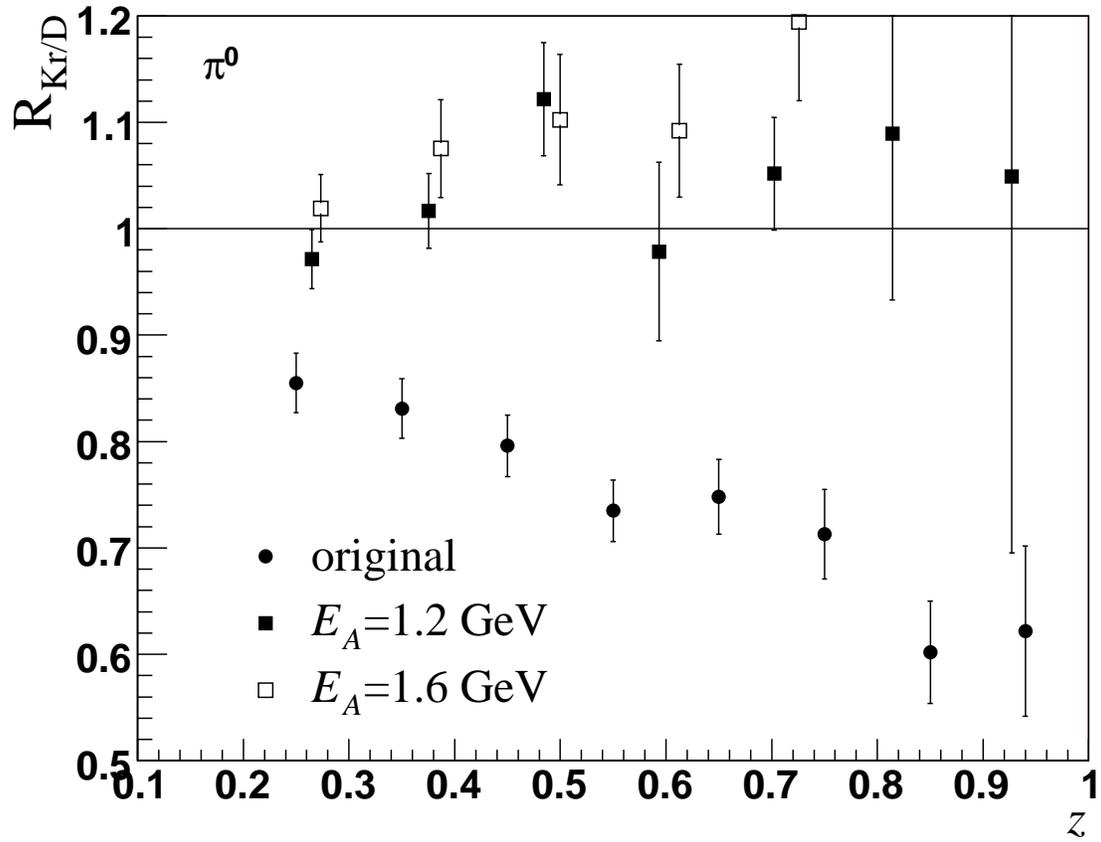}
\captionstyle{normal}
\caption{Multiplicity ratios for neutal pions from Kr target:
original ratio measured by HERMES \cite{Aira1} (solid circles), shifted
ratios
at $E_A=1.2$~GeV (solid squares) and $E_A=1.6$~GeV (open squares).}
 \label{fig:HERMES_ratios}
 \end{figure}

Now to estimate how much energy of the virtual photon is spent on
nuclear effects
one should calculate the ratio Eq.~(\ref{eq:Rhm}) of the shifted Kr
spectrum
to the hydrogen spectrum. The calculated ratios are presented in
Fig.~\ref{fig:HERMES_ratios} for different $E_A$ values (squares).
The figure also represents original ratio without any shift (solid circles).
One can see that corresponding values of energy
losses lie between $E_A=1.0$ and 1.4~GeV.

\begin{figure}[htb]
 \setcaptionmargin{5mm} 
\onelinecaptionsfalse
 \includegraphics[width=.9\hsize]{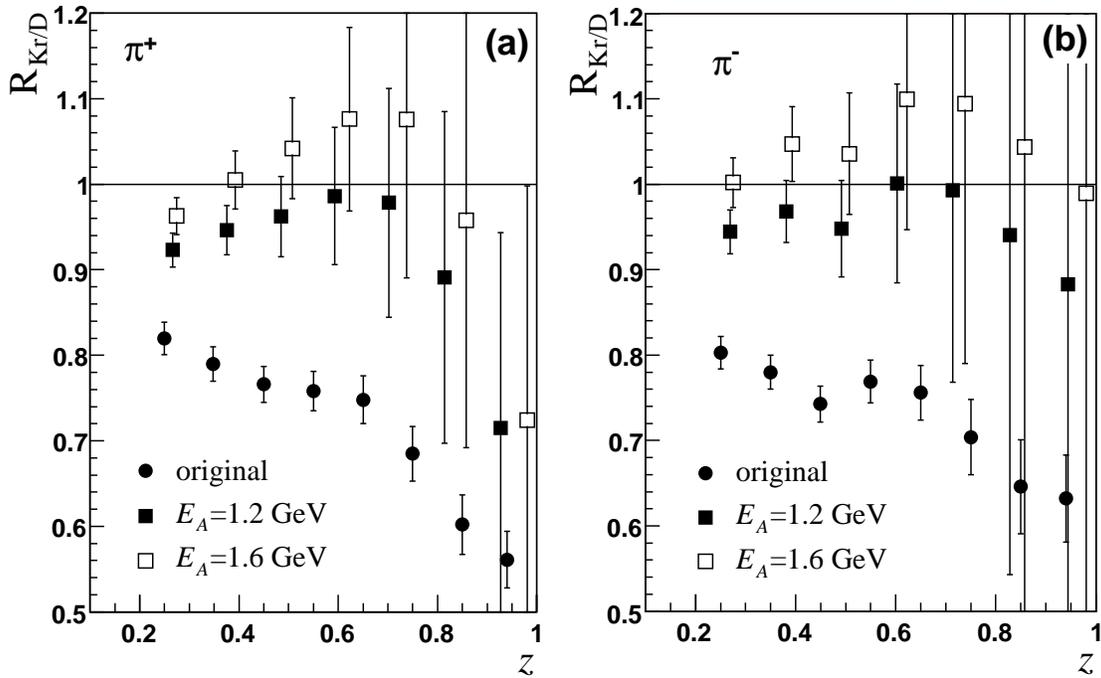}
\captionstyle{normal}
 \caption{Multiplicity ratios for (a) $\pi^+$:
original ratio measured by HERMES \cite{Aira1} (solid circles), shifted
ratios
at $E_A=1.2$~GeV (solid squares) and $E_A=1.6$~GeV (open squares);
(b) $\pi^-$: original ratio measured by HERMES \cite{Aira1} (solid
circles),
shifted ratios at $E_A=1.2$~GeV (solid squares) and $E_A=1.6$~GeV (open
squares).}
 \label{fig:HERMES_ratios_ch_pi}
 \end{figure}

Omitting intermediate calculations we present results of the analysis for
charged pions (from the same experiment) at the same energies
(see Fig.~\ref{fig:HERMES_ratios_ch_pi}a for
$\pi^+$ and Fig.~\ref{fig:HERMES_ratios_ch_pi}b for $\pi^-$). From
the last two figures (Fig.~\ref{fig:HERMES_ratios} and
\ref{fig:HERMES_ratios_ch_pi})
one can see approximately the same suppression for all pions.

\begin{figure}[htb]
 \setcaptionmargin{5mm} 
\onelinecaptionsfalse
\includegraphics[width=.9\hsize]{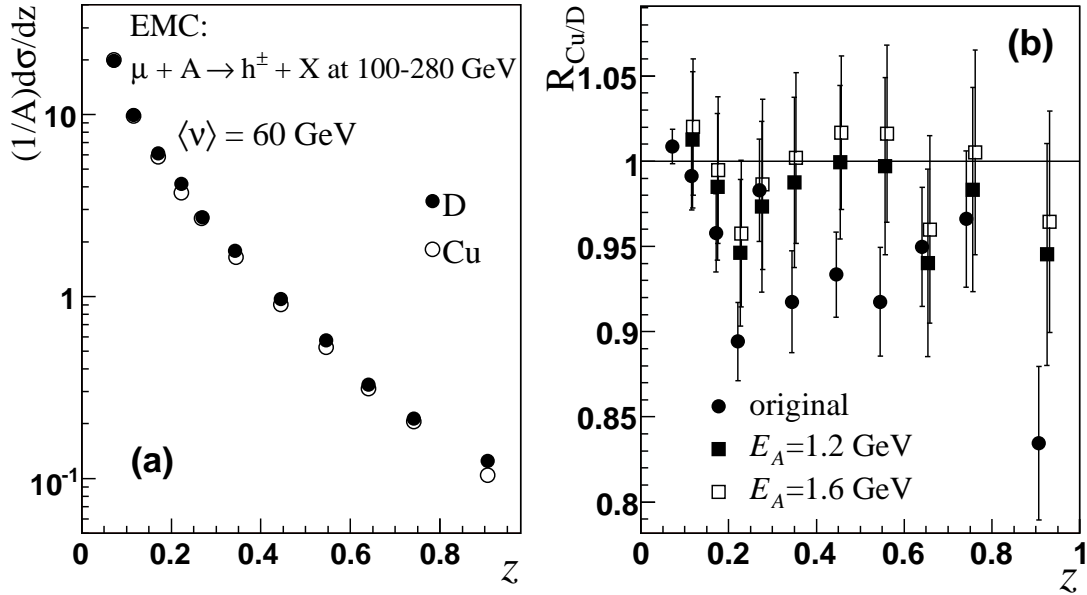}
\captionstyle{normal}
\caption{(a) Differential hadron multiplicity as a function of $z$ for Cu
(open circles) and D (solid circles). The statistical errors are of a
similar size to the symbols, the systematic errors are not shown.
(b) Multiplicity ratios for hadrons from Cu target:
original ratio measured by EMC \cite{Ashman} (solid circles),
shifted ratios at $E_A=1.2$~GeV (solid squares) and $E_A=1.6$~GeV (open
squares).}
 \label{fig:EMC}
 \end{figure}

There are additional data on the market relevant for such an analysis,
namely the data on deep-inelastic $\mu$D and $\mu$Cu scattering at
100--280~GeV
obtained by EMC \cite{Ashman}. They measured differential multiplicities of
forward produced charged hadrons on both nuclei ($\langle\nu\rangle=60$~GeV),
which can
be used directly in our analysis as described above. The results of the
analysis
one can see in Fig.~\ref{fig:EMC}. Fig.~\ref{fig:EMC}a represents
multiplicities
for D (solid circles) and Cu (open circles) targets measured by EMC, and
Fig.~\ref{fig:EMC}b represents the ratios obtained with our analysis
(squares)
as well as the original EMC ratio (solid circles).

Results for energy losses obtained for Cu target $E_A \approx 1.4$~GeV
are in reasonable agreement with those obtained for
Kr target. However, HERMES measurements were done at $\langle \nu \rangle
\approx 16$~GeV,
while EMC data were taken at $\langle \nu \rangle \approx 60$~GeV.
Therefore
one can conclude that energy  dependence of nuclear effects 
for all charged particles here is rather weak
within the errors of the analysis.

Unfortunately the inclusive spectra of identified secondaries are published only
at one (HERMES) energy and we cannot discuss the energy dependence of our
$E_A$ parameter. However, there exist the experimental results
\cite{Aira1} for $R_{A/D}$ as a function of $\nu$. For pions these ratios
increase with $\nu$ and it means that $E_A$ values have at least
 more weak $\nu$ dependence than the linear one. The values of
$R_{A/D}$ are evidently different for secondaries $K^+$ and $K^-$,
as well as for $p$ and $\bar{p}$. Principally, this difference can be
connected \cite{Bia} with different absorption cross sections of
secondaries. Another possible explanation for evident difference in
yields of secondary $p$ and $\bar{p}$ comes from the baryon charge
diffusion from the target to forward hemisphere due to string junction
mechanism \cite{Khar,Kop,ACKS,Bopp}.

\section{Difference between the secondary production in hard and soft
processes}
\label{sec:hard_soft}

Let us try to use the presented approach, see also \cite{Berd1}, for
the description of $A$ dependences of the spectra of secondaries
produced in soft hadron nucleus collisions.

Let us define the variable
\begin{equation}
\label{eq:x0}
x_0=\frac{p}{p_0} \;,
\end{equation}
where $p$ is the momentum of the produced secondary and $p_0$ is the
initial momentum of the beam particle, both in c.m. frame, and the
variable
\begin{equation}
\label{eq:xF}
x_{\rm F}=\frac{p}{p_0 - p_A} \;,
\end{equation}
where the shift $p_A$ accounts for all nuclear effects.

Let us define the ratio of the multiplicities of secondary hadrons of type $h$
produced in $h_1A$ and $h_1p$ collisions as
\begin{equation}
\label{eq:Rh}
R_{A/p}(x) = \frac{\frac{1}{A}\left(\frac{d\sigma}{dx}\right)_{h_1A\rightarrow h}}
 {\left(\frac{d\sigma}{dx}\right)_{h_1p\rightarrow h}},
\end{equation}
Similarly to Eqs.~(3) and (4) we assume that
\begin{equation}
R_{A/p}(x_{\rm F}) = const (x_{\rm F}) \simeq 1 ,
\end{equation}
whereas the same ratios in terms of $x_0$ depends on $x_0$
\begin{equation}
R_{A/p}(x_0) = f(x_0).
\end{equation}

\begin{figure}[htb]
\setcaptionmargin{5mm} 
\onelinecaptionsfalse
\includegraphics[width=.9\hsize]{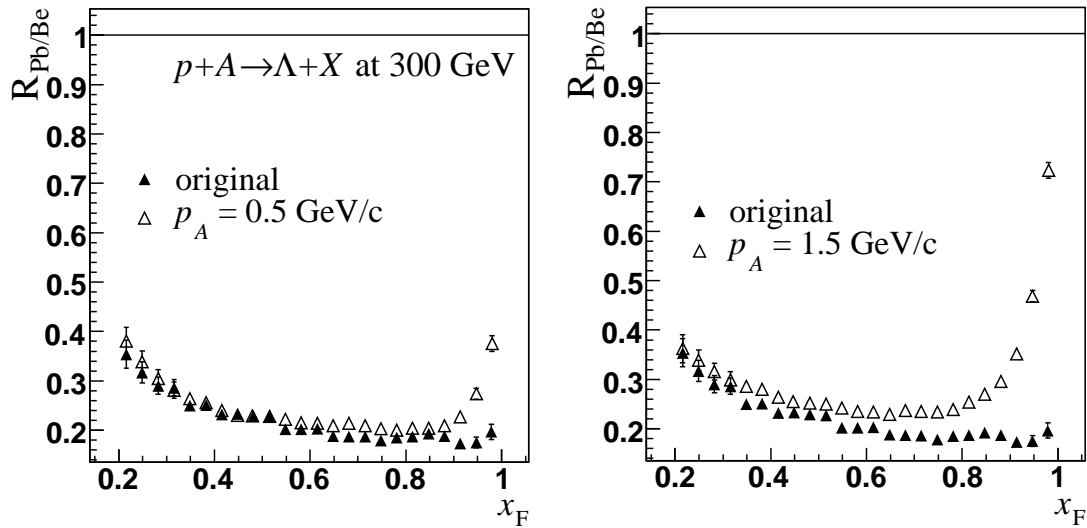}
\captionstyle{normal}
\caption{Ratios of the multiplicities of $\Lambda$ produced softly ($\theta$ = 0.25 mrad)
on Pb and Be targets at 300 GeV.}
 \label{fig:Lambda}
 \end{figure}

Such assumption leads to the agreement \cite{Berd1} with the data on
$J/\Psi$ and Drell-Yan pair production on nuclear targets with
$p_A \ll p_0$. Now we use the same assumption for the case of soft
$\Lambda$ production in $pA$ collisions, and the results for the data
\cite{Scu} are shown in Fig.~\ref{fig:Lambda}. We can see total 
disagreement with the data. The values of $R_{A/p}(x_{\rm F})$ depend 
very weakly on $p_A$ value and we can not find such $p_A$ value when 
Eq.~(9) is fulfilled.

The data for $pA \to pX$ at small $p_T$ \cite{Bart} show the behaviour 
similar to the case shown in Fig.~\ref{fig:Lambda}, whereas the 
situation with the data on soft pion production in $pA$ and $\pi A$ 
collisions \cite{Bart} is rather unclear due to the large experimental 
error bars.

We can try to explain the presented difference in nuclear effects for
hard and soft production of secondaries by the essential difference in
impact parameters which give the main contribution in these two cases.

Let us consider the $A$ dependences of the inclusive production cross
section as
\begin{equation}
x_{\rm F} d\sigma / dx_{\rm F} \sim A^{\alpha(x_{\rm F})}.
\label{eq:A_alpha}
\end{equation}

In hard collisions $\alpha(x_{\rm F}) \sim 1$ except of rather large 
$x_{\rm F}$ values. It means that the considered secondary particle can 
be produced on every target nucleon with equal probability, {\it i.e.} 
absorption effects are small. Even for the secondaries produced at large 
$x_{\rm F}$, $\alpha(x_{\rm F}) > 2/3$. So all impact parameters 
contribute here, and some decrease of $\alpha$ values at large 
$x_{\rm F}$ can be connected with the effective decrease of the beam 
energy, as it was shown in \cite{Berd1} and in Section 2 of the present 
paper.

In the case of soft production of secondaries the experimental values
of $\alpha$ at $x_{\rm F} >$ 0.2--0.3 are about 1/3 and, contrary to the 
case of hard production, they rather weakly depend on $x_{\rm F}$ in 
this $x_{\rm F}$ region. This corresponds to the picture when the 
secondaries at $x_{\rm F} >$ 0.2--0.3 are produced mainly in one-fold 
interaction of beam particle with the target nucleus. The corresponding 
cross section for $\nu$-fold inelastic interactions has the form 
\cite{Tre,Sh1}
\begin{equation}
\sigma^{(\nu)}_{prod} = \frac1{\nu !} \int d^2b [\sigma_{in} T(b)]^{\nu}
e^{-\sigma_{in} T(b)} \;,
\end{equation}
where $\sigma_{in}$ is the inelastic interaction cross section of beam 
particle with a nucleon. For hadron beam the product $\sigma_{in} T(b)$ 
is small enough only for large impact parameters. This leads immediately 
to the value $\alpha \sim 1/3$ for large $x_{\rm F}$ where multiple 
interactions ($\nu>1$) can not contribute due to the energy 
conservation. The ratio Eq.~(\ref{eq:Rh}) is determined now
by the ratio $\sigma_{prod}^{(1)}/\sigma_{prod}$ 
($\sigma_{prod}=\sum_{\nu=1}^A\sigma_{prod}^{(\nu)}$).

\begin{figure}[htb]
\setcaptionmargin{5mm} 
\onelinecaptionsfalse
\includegraphics[width=.9\hsize]{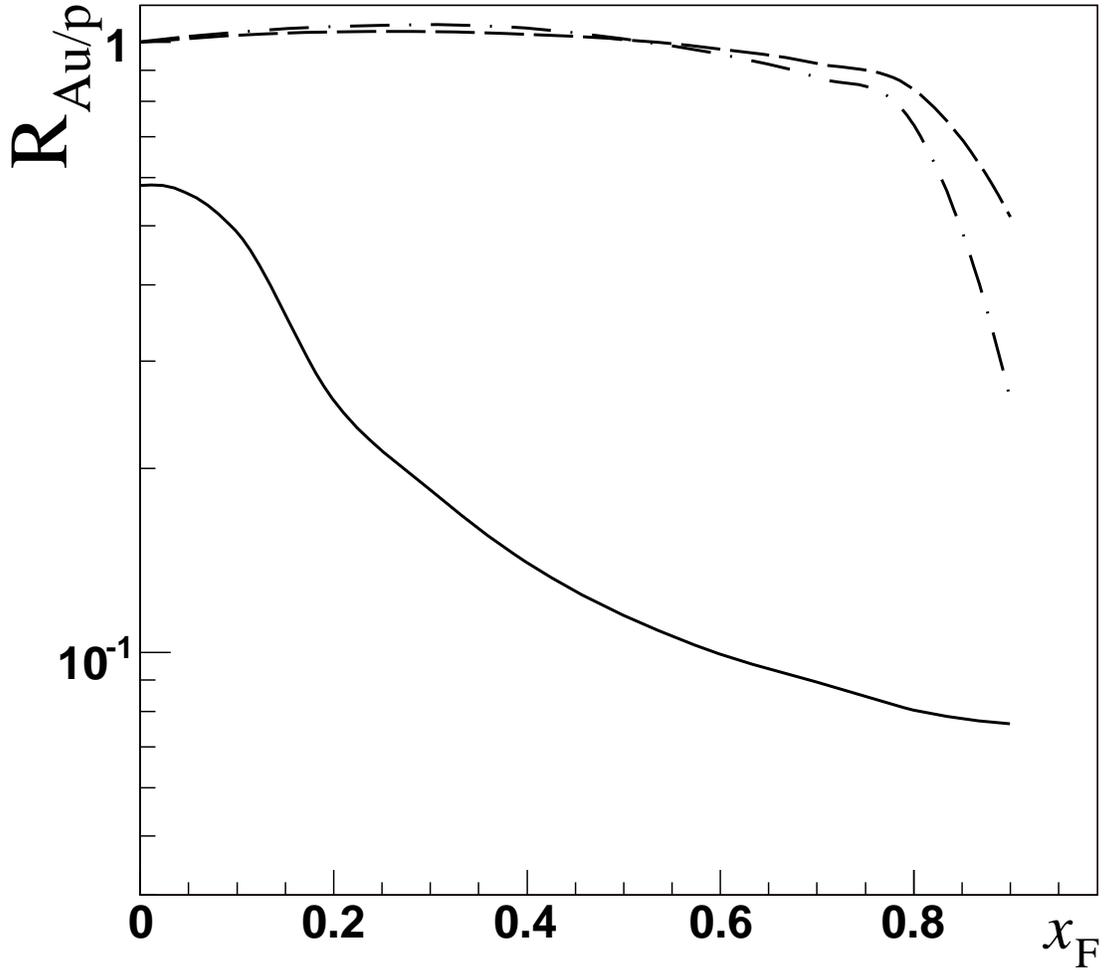}
\captionstyle{normal}
\caption{The calculated ratios of the multiplicities of $\Lambda$ 
produced in hard and soft $pA$ and $pp$ collisions at $\sqrt{s}=20$~GeV. 
The curves for ''hard'' production were obtained with $p_A=0.5$~GeV/c 
(dashed curve) and $p_A=0.75$~GeV/c (dash-dotted curve).}
 \label{fig:hard_soft}
 \end{figure}

The behaviour of $\Lambda$ yields produced in hard and soft $pA$ to $pp$ 
collisions, calculated in the framework of the Quark-Gluon String Model 
\cite{Kaidalov1,Shab} is shown in Fig.~\ref{fig:hard_soft}. The difference in 
the soft and hard interactions is reproduced rather clear. In the case 
of soft interactions we obtain rather strong $x_{\rm F}$-dependence of 
$\alpha(x_{\rm F})$ Eq.~(\ref{eq:A_alpha}) at small $x_{\rm F}$ and rather 
weak dependence at large $x_{\rm F}$. The case of ''hard production'' 
was generated with the help of rescale, Eq.~(\ref{eq:xF}), assuming 
firstly $\alpha(x_{\rm F})=1$. Here we obtain the behaviour typical for 
experimental hard production, weak $x_{\rm F}$-dependence of 
$\alpha(x_{\rm F})$ at small $x_{\rm F}$ and strong dependence at large 
$x_{\rm F}$.

\section{Conclusion}

In summary, we considered energy losses of the virtual photon in 
deep-inelastic $lA$ collisions from the available experimental data. 
What we were interested in is how much energy of the virtual photon is 
spent on all nuclear effects including the effect of nuclear 
disintegration.

Our results for energy losses $E_A\approx1.4$~GeV obtained for Kr and Cu 
targets correspond to energy loss ratios $dE/dz=$~0.5--0.6~GeV/fm (we 
assumed that DIS takes place somewhere in the center of the nucleus).
This value is $\approx 5$ times smaller than we obtained in our previous 
analysis \cite{Berd1}. However, in the present paper we deal with 
energies many times smaller than in the previous one. From comparison 
between HERMES and EMC data obtained at different virtual photon 
energies ($\nu=$~7--12~GeV and 60~GeV respectively) we conclude that our 
$E_A$ value, within the accuracy of the performed analysis,
has very weak $\nu$-dependence.

There are a lot of models for leptoproduction of hadrons in nuclear DIS. 
They use different approaches such as absorption of a prehadron state 
and modification of quark fragmentation functions in nuclei 
\cite{Accardi,APSS}, energy losses (vacuum and induced), production and 
formation times of a prehadron and a hadron respectively \cite{Kop1}. 
There is an experimental evidence for the effect of nuclear modification 
of nucleon structure functions \cite{Esk}, thus, all the models must take 
into account this effect as well. The obtained results for HERMES data 
are in reasonable agreement with the theoretical calculations of 
Refs. \cite{Kop1,Accardi}, since both cited models agree well with the 
HERMES data on pion production. Concerning our result obtained for EMC, 
in \cite{Kop1} these data ($\nu=60$~GeV) were not considered. However,
in Ref. \cite{Accardi} it was shown that the effect of absorption of the 
prehadron is negligible for these data, which implies that the main 
contribution to the observed suppression comes from the effect of
nuclear modification of quark fragmentation functions, while in the case 
of HERMES data both the effects are important.

However, concluding that some or other effect (if there is no 
experimental evidence for it) gives the main contribution to the 
considering process is questionable.  {\it E.g.} if one consideres 
models for energy loss and absorption, they both lead to a suppression 
law $\sim A^{2/3}$ (broken at A$\ge$80), thus, one cannot completely 
distinguish between these two effects \cite{Accardi2}. Actually this 
does not allow us do draw a definite conclusion which effect is dominant 
in the considered processes of nuclear DIS for HERMES and EMC. The only 
we can say is how much energy of the virtual photon is spent on 
{\it all} nuclear effects. Obviously, among the others there exist two 
effects: nuclear disintegration and modification of nuclear
structure functions (however, they are not dominant here).

In section \ref{sec:hard_soft} we have demonstrated that one could not 
apply our analysis to soft processes. We also put forward an argument 
for this fact which lies in that in the case of hard processes every 
nucleon gives an equal contribution to the secondary production cross 
section (because $d\sigma/dx\sim A^{\alpha}$, where $\alpha\sim 1$),
while in the case of soft processes only a rather small part of all the 
nucleons contribute to the cross section (because $\alpha\sim1/3$). 
Moreover, these contributing nucleons are concentrated on the 
peripherals of the nucleus. The number of these nucleons is not
changed with $x_{\rm F}$ at large $x_{\rm F}$, so the ratio of secondary 
yields on nuclear and nucleon targets is practically constant.

We are grateful to M.~G.~Ryskin for discussions.

%\newpage
%\input abstract_rus.tex
\newpage
\selectlanguage{english}

\end{document}

%% file: abstract_eng.tex
\title{Nuclear effects in leptoproduction of secondaries}

\author{\firstname{Ya.A.}~\surname{Berdnikov}}
\affiliation{St.-Petersburg State Polytechnic University, St.-Petersburg, Russia}

\author{\firstname{M.M.}~\surname{Ryzhinskiy}}
\email{mryzhinskiy@phmf.spbstu.ru}
\affiliation{St.-Petersburg State Polytechnic University, St.-Petersburg, Russia}

\author{\firstname{Yu.M.}~\surname{Shabelski}}
\email{shabelsk@thd.pnpi.spb.ru}
\affiliation{Petersburg Nuclear Physics Institute, Gatchina, Russia}

 \begin{abstract}
 We estimate the energy losses in the case of deep-inelastic scattering
on nuclear targets in terms of the effective change of the virtual
photon energy. Our phenomenological results are in reasonable agreement
with theoretical calculations. The difference in secondary production
processes in hard and soft interactions is discussed.
 \end{abstract}

%\pacs{{13.85.Qk}{}, {25.75.Dw}{}, {25.40.Ve}{}}


\maketitle